\documentclass[conference]{IEEEtran}
\IEEEoverridecommandlockouts

\usepackage{cite}
\usepackage{amsmath,amssymb,amsfonts}
\usepackage{algorithmic}
\usepackage{graphicx}
\usepackage{textcomp}
\usepackage{xcolor}
\usepackage{subcaption}
\usepackage{enumitem}
\usepackage{placeins}

\def\BibTeX{{\rm B\kern-.05em{\sc i\kern-.025em b}\kern-.08em
    T\kern-.1667em\lower.7ex\hbox{E}\kern-.125emX}}
\begin{document}

\title{\huge KPI Poisoning: An Attack in Open RAN Near Real-Time Control Loop}

\author{\IEEEauthorblockN{Hamed Alimohammadi, Sotiris Chatzimiltis, Samara Mayhoub, \\ Mohammad Shojafar, Seyed Ahmad Soleymani, Ayhan Akbas,
Chuan Heng Foh}
\IEEEauthorblockA{5G/6G Innovation Centre, Institute for Communication Systems, University of Surrey, Guildford, UK}
Email: { \{h.alimohammadi, sc02449, s.mayhoub, m.shojafar, s.soleymani, a.akbas, c.foh\}@surrey.ac.uk}}

\maketitle

\begin{abstract}
Open Radio Access Network (Open RAN) is a new paradigm to provide fundamental features for supporting next-generation mobile networks. Disaggregation, virtualisation, closed-loop data-driven control, and open interfaces bring flexibility and interoperability to the network deployment. However, these features also create a new surface for security threats. In this paper, we introduce Key Performance Indicators (KPIs) poisoning attack in Near Real-Time control loops as a new form of threat that can have significant effects on the Open RAN functionality. This threat can arise from traffic spoofing on the E2 interface or compromised E2 nodes. The role of KPIs is explored in the use cases of Near Real-Time control loops. Then, the potential impacts of the attack are analysed. An ML-based approach is proposed to detect poisoned KPI values before using them in control loops. Emulations are conducted to generate KPI reports and inject anomalies into the values. A Long Short-Term Memory (LSTM) neural network model is used to detect anomalies. The results show that more amplified injected values are more accessible to detect, and using more report sequences leads to better performance in anomaly detection, with detection rates improving from 62\% to 99\%.

\end{abstract}


\section{Introduction}
The complexity of wireless cellular network management has significantly increased with new technologies, such as massive Multiple-Input Multiple-Output (mMIMO) and terahertz communications. Maintaining and keeping this heterogeneous environment running is more expensive for Mobile Network Operators (MNOs). Traditional all-in-one solutions are not capable of supporting the network's configuration requirements. Recently, the idea of opening up the Radio Access Network (RAN), inherited from Software-Defined Networking (SDN), has been brought to the attention as a solution for the mentioned problem. Open RAN is a new paradigm born with this idea. It provides fine-grained configuration/reconfiguration and closed-loop control and removes vendor lock-in to make it possible to use multi-vendor heterogeneous components inside a network. Virtualisation, disaggregation, closed-loop data-driven control, open interfaces, and Artificial Intelligence/Machine Learning (AI/ML) facilities are the most critical features of Open-RAN, which enables it to overcome the current challenges of traditional RANs by presenting more flexibility and cost-effectiveness~\cite{bib-1}.

The most common disaggregation model for a base station used by Open-RAN architectures is 3GPP NR 7.2 split~\cite{bib-2}. It disaggregates a base station into a Central Unit (CU), a Distributed Unit (DU), and a Radio Unit (RU). Moreover, CU is divided into User Plane (UP-CU) and Control Plane (CP-CU). In Open-RAN, these components are called Open-CU (O-CU), Open-DU (O-DU), and Open-RU (O-RU) to indicate the openness and interoperability of the components in the Open-RAN environment.

\begin{figure}[htbp]
\centerline{\includegraphics[width=\columnwidth]{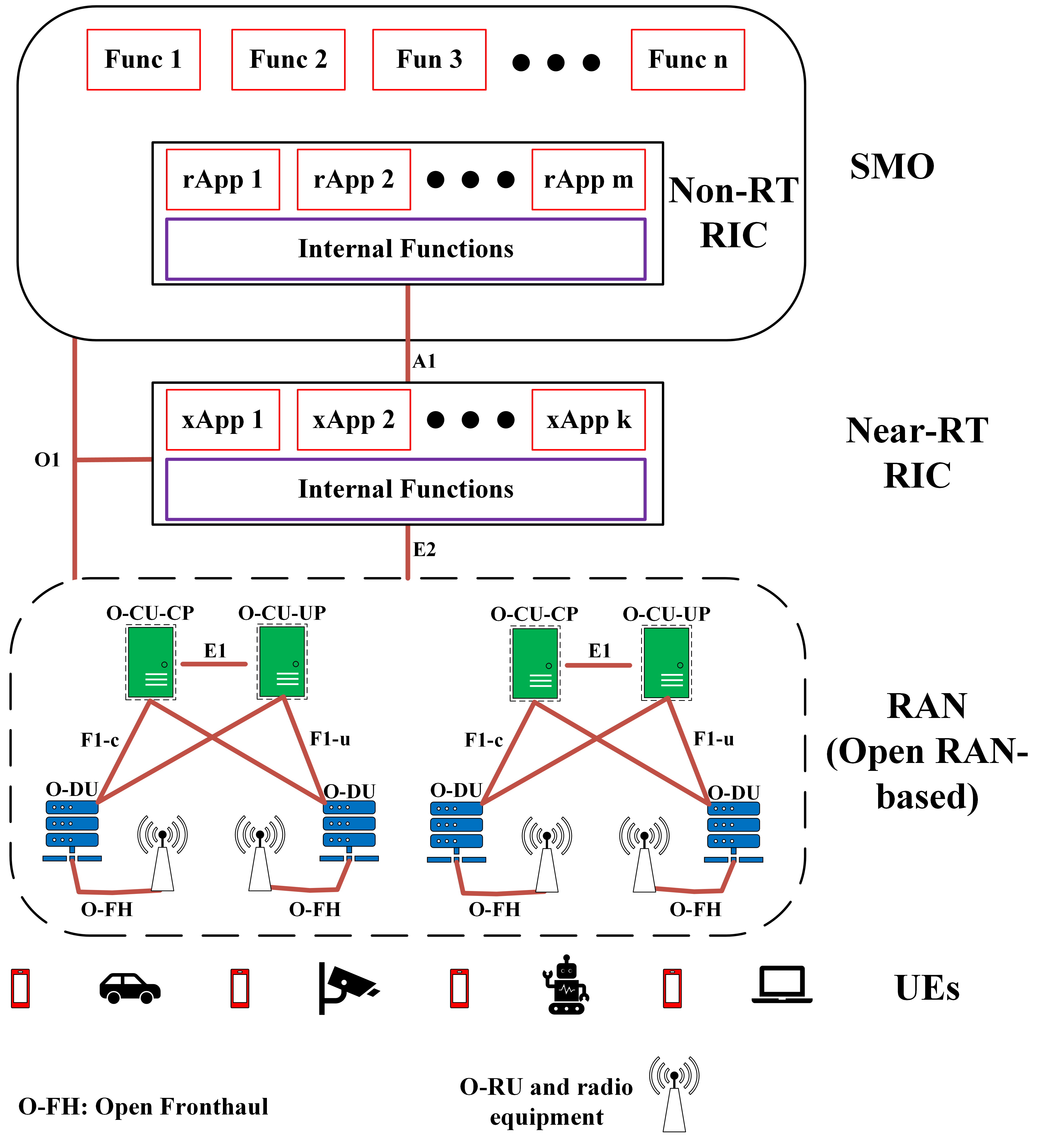}}
\caption{\small Open RAN overall architecture.}
\label{ORAN}
\end{figure}

Also, RAN Intelligent Controllers (RIC) are new components connected to the mentioned units through open interfaces and monitor and control the network in closed control loops. RIC itself is separated into two parts with different control time scales. Near Real-Time RIC (Near-RT RIC) control loop time scale is between 10\textit{ms} to 1000\textit{ms}, and the control loops with more than 1000\textit{ms} time scale are done by Non-Real Time RIC (Non-RT RIC). It should be noted that Non-RT RIC resides in the Service Management and Orchestration (SMO) platform, which manages and orchestrates network resources and services. Additionally, third-party applications perform monitoring and controlling tasks on top of Near-RT and Non-RT RICs. The applications running on Near-RT RIC and Non-RT RIC are called xApp and rApp, respectively. These applications can carry out various missions, such as monitoring and resource allocation. They can subscribe to receive KPI reports from RAN components, including information such as throughput, delay, number of physical resource blocks used, channel measurements, and more. These reports are utilised for monitoring and decision-making when issuing control actions to the network.
Finally, the open interfaces, called A1 and E2, connect Non-RT RIC to Near-RT RIC and Near-RT RIC to the RAN components, respectively. O-CUs and O-DUs are connected to Near-RT RIC through E2 interfaces; therefore, they can be called E2 nodes. The other interfaces and overall architecture of Open RAN can be seen in Fig.~\ref{ORAN}. 

\subsection{The main Contribution of this paper}
This paper introduces an attack that can severely damage the network. As mentioned above, KPI reports control the network, so an attacker may change the values in the reports to impose wrong decisions on the controllers. We call this attack \textit{KPI Poisoning Attack}. Its potential impacts are analysed and an ML-based approach is proposed to detect it. LSTM neural networks is used as the core model of our detection method. The scope of this paper is Near-RT RIC level control loops. Note that the service model proposed by the O-RAN Alliance, called E2SM, is used to report specific measurements, known as Key Performance Measurements (KPMs), from E2 nodes to the Near-RT RIC~\cite{bib-5}. In this paper, we use the KPI term to be more general, as measurements can be collected through several standards and services. Our contribution can be summarized as:
\begin{itemize}
    \item Exploring KPIs' role in different use cases of Near-RT RIC.
    \item Introducing a threat, KPI Poisoning Attack, in Near-RT RIC control loops.
    \item Impact analysis of the KPI Poisoning Attack in the mentioned use cases.
    \item Proposing an ML-based approach, including a workflow architecture, to detect/mitigate anomalies in KPI values before being utilised in Near-RT RIC control loops.
\end{itemize}

The remainder of the paper is organised as follows.
Section~\ref{RelatedWorks} reviews related work.  Section~\ref{main-1} discusses using KPIs in Open RAN Near-RT RIC. KPI Poisoning attack and our detection approach are described in section~\ref{main-2}. Experimental results are presented in section~\ref{experiments}. Finally, the paper is concluded in section~\ref{conclusion}.

\section{Related Work} \label{RelatedWorks}

Recent works in the literature describe the security challenges of the Open RAN architecture in general. For example, the papers~\cite{bib-3},~\cite{bib-7},~\cite{bib-8},~\cite{bib-9}, and~\cite{bib-10} detail the security risks associated with the key aspects of Open RAN, such as its intelligence, open interfaces, virtualisation, and disaggregation. They also discussed potential threats against AI/ML components in Open RAN (xApp/rApp). These threats include: i) Data poisoning attacks - when an attacker with unauthorized access poisons the training dataset. ii) Evasion attacks - when an attacker uses adversarial input data to force AI/ML models to make mistakes and degrade the network performance~\cite{bib-8}. In addition, previous studies~\cite{bib-9} and~\cite{bib-10} have theoretically examined the sources of attacks in Open RAN, mainly through the E2 interface. It includes Denial-of-Service (DoS), Man-in-the-Middle (MitM), protocol exploitation attacks on E2AP, unauthorized access, replay attacks, and manipulation and tampering of E2 data (as in our work). The studies also investigated potential threat agents against Near-RT RIC and the E2 interface, such as compromised or malicious xApps, compromised or malicious E2 nodes, and external agents via open interfaces (e.g. E2, A1, and O1).

In addition, the security threats of integrating AI/ML into Near-RT RIC were discussed in~\cite{bib-11}, focusing on malicious xApps as threat agents. The study exploits the access to other components of the Near-RT RIC, such as the shared database, RIC Message Router, conflict manager and E2 manager. The studied attacks included data and model poisoning, evasion attacks, membership inference attacks and model extraction attacks. However, the authors did not focus on any other threat agents (such as malicious E2 nodes) and did not conduct any experimental studies. 

Furthermore, the paper~\cite{bib-12} summarized the O-RAN Alliance WG11 ~\cite{bib-20} and ~\cite{bib-21} threat models against the A1, E2 and F1 open interfaces and conducted experimental attacks against them. One experiment studied the influence of the exploited E2 interface on two xApps: i) KPI monitoring xApp and ii) Network slicing xApp. The authors applied adversarial effects (delay and loss) to the E2 interface. Then they studied the impact on the performance of the mentioned xApps. However, they did not study the case of malicious KPI values being injected.

There are works explicitly related to our study on KPI Poisoning Attack. For instance, in~\cite{bib-13}, the authors mention the new vulnerabilities introduced against the intelligent components of Open RAN brought by the open interfaces. The open interfaces could be exploited by MiTM, which could inject malicious KPI reports into the E2 interface toward the Near-RT RIC or malicious control actions from the Near-RT RIC toward the E2 nodes. The paper~\cite{bib-13} demonstrated adversarial machine learning attacks targeting the input KPI reports of network slicing xApp from ~\cite{bib-14}, implemented using Deep Reinforcement Learning (DRL) model. The authors introduced a mitigation method using Autoencoders (AEs) to address these attacks. In their study, they tampered with the KPI values by adding random noise following a normal distribution.

Additionally, in the study by paper~\cite{bib-15}, the impact of a malicious xApp manipulating input data in the database in the Near-RT RIC was examined. The authors focused on a use case involving a Deep Neural Networks (DNN)-based interference classification xApp trained using the public 5G Radio Frequency Interference (RFI) dataset~\cite{bib-16}. Two techniques were employed to generate adversary input: i) the Fast Gradient Sign Method (FGSM)~\cite{bib-17} and ii) the Projected Gradient Descent (PGD) attack~\cite{bib-18}. Notably, this paper did not utilise KPI reports as a data source for their xApp. Similarly, the authors in~\cite{bib-19} developed an attack model targeting two types of xApps: a CNN-based Interference Classification xApp (using spectrogram data) and a DNN-based Interference Classification xApp (using KPM data from KPI reports). They employed PGD and FGSM methods to generate malicious data and proposed a distillation mitigation technique.

Several studies have analysed the security threats to Open RAN, with some focusing on attacks against AI/ML models. While some researches have proposed mitigation strategies by making their models tolerable against tampered KPI values and compared xApp performance before and after such attacks, none have addressed anomaly detection, regardless of the attack source (e.g., E2 nodes, malicious xApps, E2 interfaces).
Since KPIs are crucial for AI/ML training data in Open RAN, tampering can impact the performance of xApps/rApps during both the training and inference phases. This work focuses on the impact of KPI tampering during the inference phase, specifically, in Near-RT RIC control loops by analysing the impacts in diverse use cases. We developed an ML-based anomaly detection scheme, which includes a workflow architecture and an LSTM model to identify tampered KPI values.

\section{KPIs in Open-RAN Near Real-Time Control Loop}\label{main-1}
KPI reports are crucial in RAN control loops for monitoring, generating policies, issuing control actions, and determining how RAN must work. As mentioned before, the scope of this paper is Near-RT RIC control loops. Therefore, we have chosen the Use Cases defined by Work Group 3 in~\cite{bib-6} as our baseline. The use cases are briefly explained, which are clear instances of KPIs used in Open RAN Near Real-RT RIC, to give more insight into the importance of KPI reports. 
\begin{itemize}
    \item \textit{Use Case 1 - Traffic Steering:} It uses performance metrics and network status for ML-based prediction and makes decisions on User Equipment (UE) to cell association, handover, Carrier Aggregation, and dual connectivity management. Carrier aggregation uses multiple frequency bands of channels to provide a higher data rate. Dual connectivity connects a UE to two base stations, mainly one for control procedures and one for data transmission.
    \item  \textit{Use Case 2 - QoS-based Resource Optimisation:} It is used such that resource allocation is efficient and meets QoS requirements in terms of reliability, delay, bandwidth, etc. In this use case, Near-RT RIC uses measurement reports from E2 nodes for ML-based prediction and decides on generating resource allocation policies and commands.
    \item \textit{Use Case 3 - RAN Slice SLA Assurance:} Service Level Agreement (SLA) between the mobile operator and business customer defines the capabilities to support the service delivered to the customer. These capabilities can be in various specifications, such as data rate, traffic capacity, user density, latency, reliability, and availability. Network slicing is a crucial feature to enable SLAs. This is a very challenging problem, and open interfaces and AI/ML support of Open RAN architecture can effectively solve it. Near-RT RIC uses the UE and slice level measurements to monitor and predict, and if needed generate policies and control actions to send to E2 nodes. These policies and commands are generated to ensure slice SLA indicators are not violated. It can be done by resource reallocation or preventing some UEs from using a specific slice.
    
    \item \textit{Use Case 4 - Massive MIMO Optimisation:} Massive MIMO is a technology where base stations or access points are equipped with hundreds or thousands of antennas. This use case considers the problem of Non-GoB BF (Non-Group of Beams BeamForming) as massive MIMO optimisation. Beamforming is a technique used in wireless communication to direct the transmission of radio signals in specific directions rather than broadcasting them uniformly in all directions. This improves signal strength, reduces interference, and increases overall network efficiency. Near-RT RIC uses measurement reports from E2 nodes along with other RAN configurations to train and deploy the ML model to perform the mentioned task.
    
    \item \textit{Use Case 5 - QoE Optimisation:}  As QoS can not respond to all requirements of new applications like Virtual Reality (VR), there is a need for some procedures to monitor and maintain the Quality of Experience (QoE). To this end, RAN Analytics Information (RAI) can be inferred by ML models that use measurement data sent by E2 nodes. In this use case, Near-RT RIC provides RAI through QoE-related ML models per UE or a group of UEs, e.g. per slice or cell. The RAI analytics are used by RAI service consumers, which can be an application that needs to know the QoE level of its users, such as a VR application.
    \item \textit{Use Case 6 - Network Energy Saving:} Energy consumption can be reduced in different time scales with various mechanisms, including carrier and cell switch on/off, radio frequency reconfiguration, and advanced sleep modes. To this end, Near RT RIC collects energy-saving related measurements and resource usage of the UEs. It uses this information to find solutions to optimise energy consumption using ML models. This use case works closely with traffic steering to direct traffic, such that energy usage efficiency is considered a target in ML models.
\end{itemize}

 Considering these use cases, it is clear that KPI reports directly impact operationality efficiency and performance targets, such as Quality of Service (QoS), Quality of Experience (QoE), and energy consumption. As a toy example, we further describe a QoS-based resource allocation xApp to clarify the role of KPI reports in RAN control. The xApp may decide to use more bandwidth for a UE if it finds out the expected QoS level is not being provided for that UE. This may result in higher power consumption in the connected base station. The same xApp may reduce assigned bandwidth if the UE transmits a low data rate. As a simple example, assume that the resource allocation xApp assigns one  Physical Resource Block (PRB) for each \textit{X Mbps} of the sum of downlink (DL) and uplink (UL) throughputs for a UE. One PRB can be a combination of frequency bands and time slots. If the monitored throughput of the UE is \textit{Y}*\textit{X Mbps} less than expected, the xApp will assign \textit{Y} more PRBs to it. In the reverse way, if the xApp detects that the currently being used PRBs are less than assigned, it will appropriately decrease the number of PRBs for that UE. 

\section{KPI Poisoning Attack and a Detection Approach}
\label{main-2}

Considering the crucial role of KPIs in Open RAN, it is undoubtedly one of the most probable targets for attackers to poison the KPI reports by manipulating the values. It can be random changes in the values or consciously manipulating some values to make some planned consequences. In this paper, we consider random changes. It is assumed that Near-RT RIC is fully trusted and poisoning occurs before the reports arrive. Therefore, Poisoning KPIs can be achieved in two scenarios: 1) MiTM attacks on E2 interfaces that spoof reports transmitted from E2 nodes to Near-RT RIC by exploiting open interface vulnerabilities or 2) compromised E2 nodes
controlled by attackers that send false reports. In this section, the attack's impact on explained use cases in section~\ref{main-1} are analysed. Then, propose a detection method.

\subsection{Impact Analysis}
Here, the impact of poisoned KPI values in each use case is considered to explore the potential consequences that it can have.

\begin{itemize}
    \item \textit{Use Case 1 - Traffic Steering}: False KPI values will lead to false prediction, mainly improper handover and carrier aggregation. For example, wrong handovers can be triggered by tampered throughput values. It can result in loss of connection quality for many UEs. It can cause severe damage to QoS, QoE, and even the ability of the network to provide a service. 
    \item \textit{Use Case 2 - QoS-based Resource Optimisation}: The clear impact of false KPIs in this use case is to miss the QoS targets. In practice, the network management systems calculate QoS scores at different levels, which show the percentage that a targeted QoS requirement is met. Therefore, we can say that the attack will reduce these scores. If Near-RT RIC is fooled by low PRB usage of UEs from a high-priority service, it may decrease the allocated PRBs, which will cause considerable damage to the service and its applications. 
    \item \textit{Use Case 3 - RAN Slice SLA Assurance}: Manipulated KPI values can change the prediction output of Near-RT RIC used for RAN control decision-making to ensure SLA in each slice. Therefore, it will result in SLA violations.  
    \item \textit{Use Case 4 - Massive MIMO Optimisation}: Fooling the prediction model of this use case can dramatically reduce the quality of signals that a UE receives, as it determines the beamforming decisions.
    \item \textit{Use Case 5 - QoE Optimisation}: The KPI poisoning can lead to wrong RAI analytics; therefore, the RAI service users will make wrong configurations or adjustments in their application. Finally, it will cause poor QoE in the target applications of RAI service consumers. Here, there can be scores for QoE, as well.
    \item \textit{Use Case 6 - Network Energy Saving}: The KPI poisoning attack can severely damage this use case. For example, if Near-RT RIC is fooled such that it decides to switch a group of cells off to save energy, it will lose the connection of many UEs, or at least unnecessary handovers.

\end{itemize}

According to the potential impacts that a KPI poisoning attack can impose on the Near-RT control loop, it is clear that in all the use cases, the operator's reputation, financial status, and worth are damaged. In Table~\ref{tab-1}, the impact analysis is summarised and the severity of the attack in the use cases is classified to three levels, low, moderate, and high. We do not take the application type into account in this severity classification. Therefore, there may be a vital application that the attack in all use cases can have severe consequences. Here, only the regular mobile network services are considered. In Use Case 1, due to the critical role of traffic steering, it has been classified as High. It is a complicated task with many controlling aspects, such as UE to cell association, network slice management, resource allocation, etc. We considered the severity of use cases 2-4 as moderate because the impact will be on the service's quality, which will impact the operator's reputation and lead to financial loss due to the drop in the company's worth. Use Case 5, QoE optimisation will impact the experience of just a group of UEs using specific applications that are RAI service consumers. In Use Case 6, the attack may result in loss of coverage in some areas, a type of network outage. Therefore, it is considered a high-impact use case.

\begin{table}[!t]
\centering
\caption{\small KPI poisoning Impact and Severity}
\label{tab-1}
\begin{tabular}{|c|p{2.8cm}|p{2.8cm}|c|}
\hline
\textbf{\#} & \textbf{Use Case}                  & \textbf{Main impact field}                    & \textbf{Severity} \\ \hline
1  & Traffic Steering                  & Handovers                           & High \\ \hline
2  & QoS-based   Resource Optimisation & QoS scores                          & Moderate \\ \hline
3  & RAN Slice SLA   Assurance         & SLA   requirements                  & Moderate \\ \hline
4  & Massive MIMO   Optimisation       & Signalling   quality                & Moderate \\ \hline
5  & QoE   Optimisation                & QoE scores                          & Low      \\ \hline
6  & Network   Energy Saving           & Cells, Radio-level   configurations & High     \\ \hline
\end{tabular}
\end{table}

\subsection{ML-based Detection Approach}
\label{detection}
Our proposed approach aims to develop an anomaly-based detection system to identify and differentiate between legitimate and falsified KPIs, ensuring the integrity of data consumed by the Near-RT RIC and xApps for policy enforcement and control actions. We employ LSTM neural networks due to their suitability for sequential datasets, as the KPI reports are created and sent in time series. Each row in a report that belongs to a specific time and user is called a record. The developed LSTM model utilises different numbers of records as inputs, to detect anomalies in the KPIs provided by the E2 nodes over time. 

\begin{figure}[htbp]
\centerline{\includegraphics[width=\columnwidth]{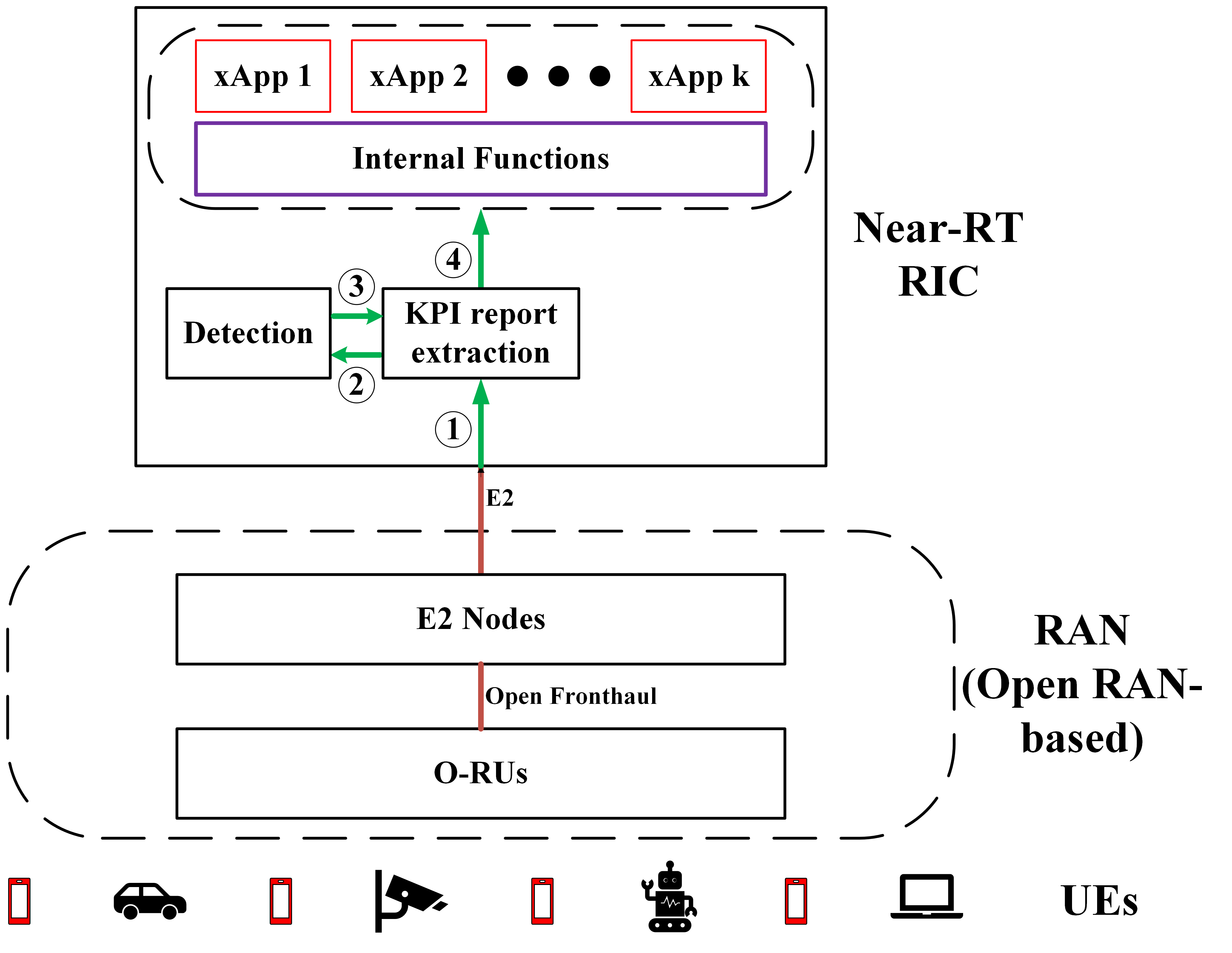}}
\caption{\small Detection/Mitigation architecture.}
\label{detect}
\end{figure}

The process begins with KPIs generated by the E2 nodes, transmitted through the E2 interface to the Near-RT RIC. Upon reception, the detection system processes these KPIs to identify potential anomalies. If the KPIs are deemed suspicious, they are tagged before being fed to the subscribed xApps. This prevents unreliable data consumption, leading to the stated impacts in the previous subsection. The proposed architecture for this mitigation approach is shown in Fig.~\ref{detect}. A KPI report extraction unit has been defined, and the received messages are inspected to check whether each message contains KPI reports. The stages of the workflow can be summarised as:
\begin{itemize}
\item Stage.1: A message is received. 
\item Stage.2: If the message includes KPI reports, it is passed to the detection unit. 
\item Stage.3: The detection unit checks the reports for poisoned KPIs. The outcome is returned to the KPI report extraction unit.
\item Stage.4: The message is tagged as benign or poisoned and passed to the functions and xApps. 
\end{itemize}

 Therefore, the functions and xApps can decide how to handle the situation, which is out of the scope of this paper. One policy can be to discard the poisoned reports and send an attack flag-up notification to SMO to manage the issue. 

\section{Performance Evaluations}
\label{experiments}
In this section, the emulation setup is explained first, followed by the presentation and description of the results.
\begin{table}[!t]
\centering
\caption{\small Dataset Features}
\label{tab-2}
\begin{tabular}{|c|p{2cm}|p{5cm}|}
\hline
\textbf{\#} & \textbf{Feature Name} & \textbf{Description}                                          \\ \hline
1           & Timestamp             & The date and time when the data was recorded.               \\ \hline
2           & UEid                  &  The unique identifier for the user equipment                                      \\ \hline
3           & UEThpUl               & User equipment uplink throughput \\ \hline
4           & PrbUsedUl             & Number of uplink physical resource blocks used  \\ \hline
5           & UEThpDl               & User equipment downlink throughput   \\ \hline
6           & PrbUsedDl             & Number of downlink physical resource blocks used \\ \hline
7           & TotNbrUl\_per\_sec    & The total number of uplink data packets per second          \\ \hline
8           & TotNbrDl\_per\_sec    & The total number of downlink data packets per second        \\ \hline
\end{tabular}
\end{table}
\vspace{-3px}
\subsection{Setup and Scenarios}
To investigate the problem of false KPI injection by MiTM attack or compromised E2 nodes, we conducted an emulation using a commercial Open RAN emulator involving 50 UE devices and three base stations. Each base station consisted of one O-CU and three O-DUs. The emulation spanned 3 hours, during which data was collected from each UE every second of its connection to the network.

The UEs were evenly distributed between two network slices: 25 UEs were assigned to the Enhanced Mobile Broadband (eMBB) slice, and 25 UEs to the Ultra-Reliable and Low Latency Communications (URLLC) slice. For each UE, six key performance indicators (KPIs) were recorded based on the capabilities of our emulation infrastructure. These six KPIs and the timestamp and UE identifier served as features for developing an LSTM-based detection model. After the emulation, the collected reports were used to train and test the LSTM model. The specifics of these features are detailed in Table~\ref{tab-2}.

A subset of UEs from each slice was randomly selected to simulate false KPI injection, with their KPIs manipulated during randomly chosen timeframes. The injection process involved calculating the multivariate normal distribution for each selected UE and amplifying its mean and covariance matrices by a factor of 1.2 to 1.5. The legitimate KPIs were replaced with these amplified values, which were then transmitted to the Near-RT RIC as usual. This scenario presents the challenge of distinguishing between genuine and synthetic KPIs provided by the E2 nodes.

The experiments were conducted on a computer with an Intel(R) Core(TM) i7-8700 CPU running at 3.20GHz and 16GB of RAM. A neural network has been implemented using the TensorFlow framework, featuring three LSTM layers with 256, 128, and 64 units, each followed by a 20\% dropout layer to capture complex temporal dependencies and reduce overfitting effectively. The model was compiled with the Adam optimiser at a learning rate 0.001 and trained using categorical cross-entropy as the loss function to classify the time series data.

\subsection{Results and discussion}
\begin{figure}[!tpb]
    \centering    \includegraphics[scale=0.5]{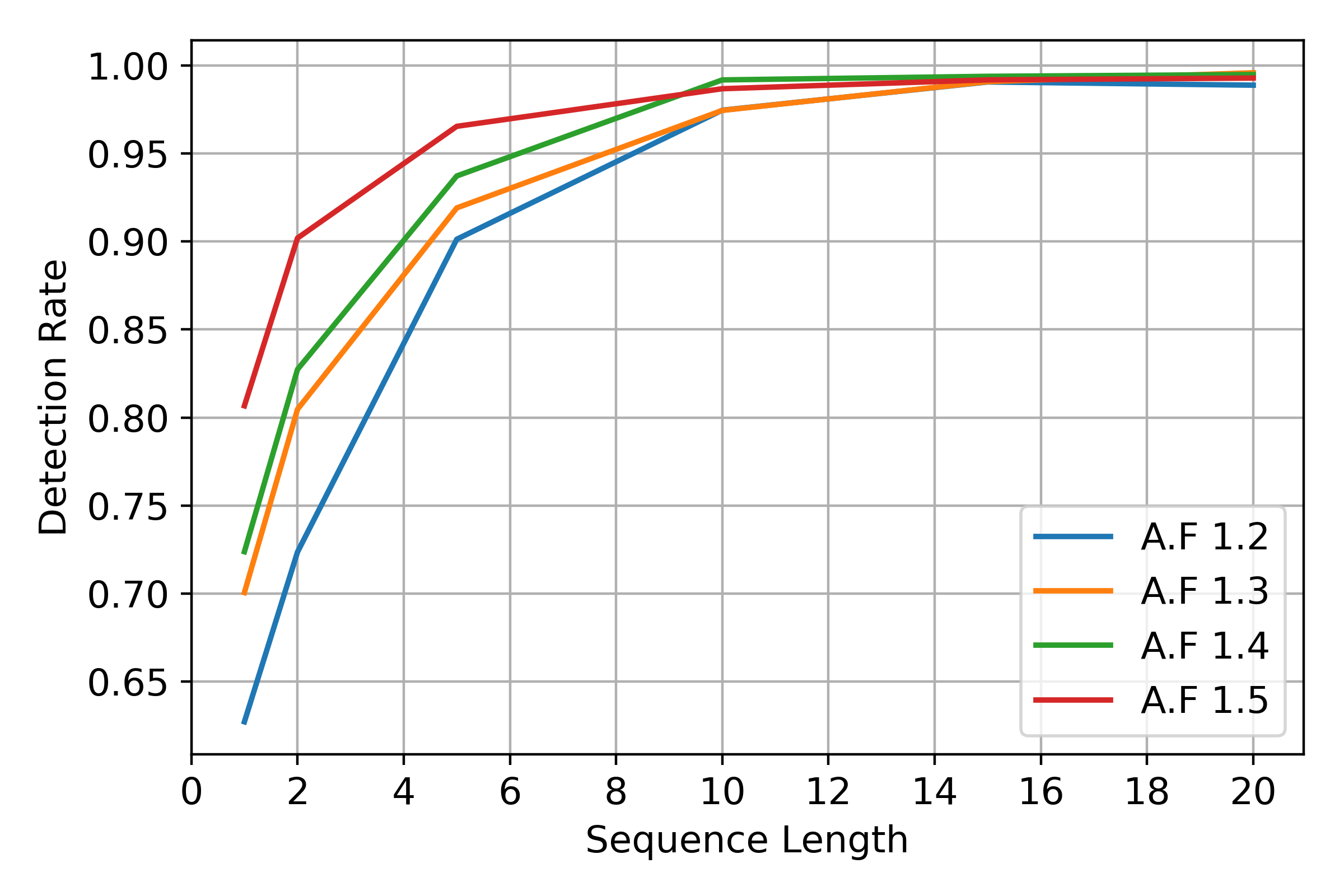}
        \caption{\small \small DR as a function of sequence length for different amplification factors.}
    \label{fig:dr}
\end{figure}

\begin{figure}[!tpb]
    \centering    \includegraphics[scale=0.5]{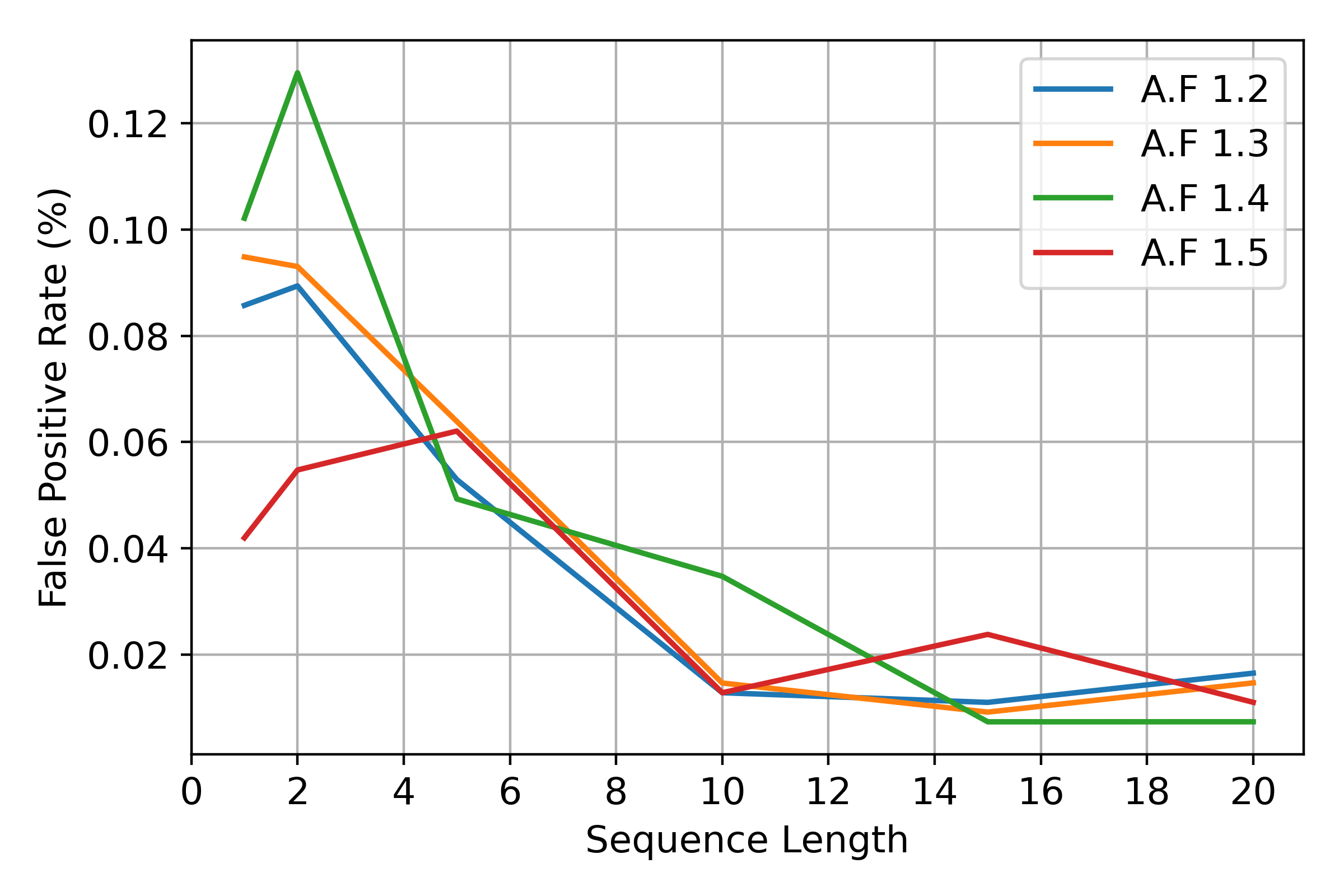}
    \caption{\small \small FPR(\%) as a function of sequence length for different amplification factors.}
    \label{fig:fpr}
\end{figure}

\begin{figure}[!tpb]
    \centering    \includegraphics[scale=0.5]{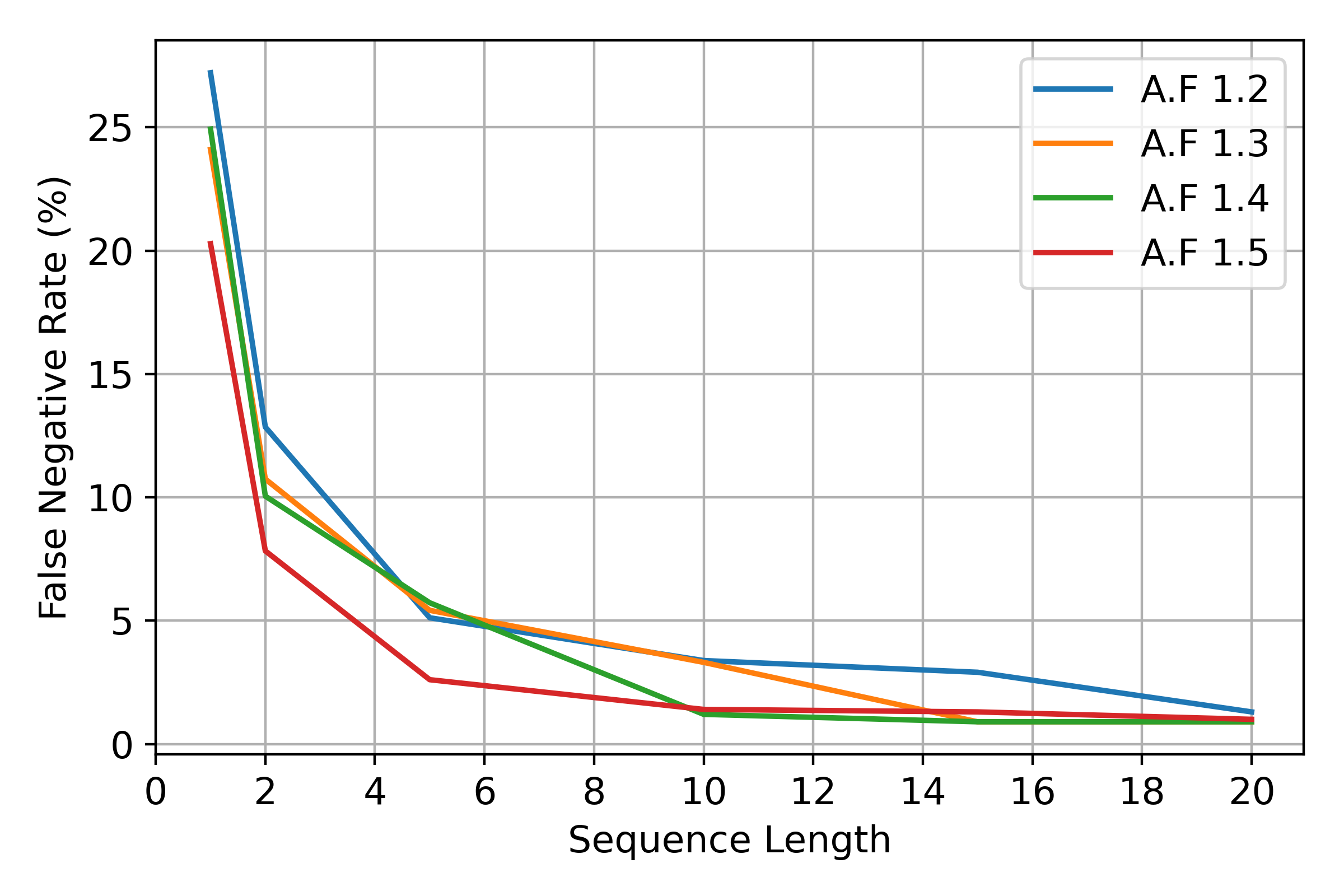}
    \caption{\small \small FNR(\%) as a function of sequence length for different amplification factors. \vspace{-10px}}
    \label{fig:fnr}
\end{figure}

The findings are illustrated in Fig.~\ref{fig:dr}, Fig.~\ref{fig:fpr}, and Fig.~\ref{fig:fnr} and are summarized as follows. Increasing the sequence length improves the detection system's performance by raising the Detection Rate (DR) and reducing both the False Positive Rate (FPR) and False Negative Rate (FNR). This suggests that more data points enhance the system's ability to distinguish between normal and synthetic data. The figures indicate that data crafted with higher amplification factors are detected more effectively. In contrast, data created with lower amplification factors are more likely to be misclassified, leading to higher FPRs and FNRs. These findings highlight the detection system's sensitivity to the synthetic data's characteristics, particularly the degree of deviation from normal patterns. Synthetic data that deviate significantly from the normal sequences are more accessible for the system to detect, likely due to the more obvious differences in the underlying patterns. Therefore, when considering strategies to synthesize data that might evade detection, employing lower amplification factors is a more effective approach, as it reduces the detectability of the synthetic data, but it will probably reduce the impact of the attack, as well. However, it is important to note that while lower amplification factors might reduce the likelihood of detection, they also require careful consideration of sequence length, as shorter sequences are more prone to misclassification errors. This leads to DR ranging from 62\% to 99\%. Thus, these results suggest a combination of sequence length and amplification factor that must be carefully balanced to optimise both the detection and evasion strategies.

\begin{figure*}[h] 
    \centering
    \begin{minipage}{0.3\textwidth}
        \centering
        \includegraphics[width=0.8\textwidth]{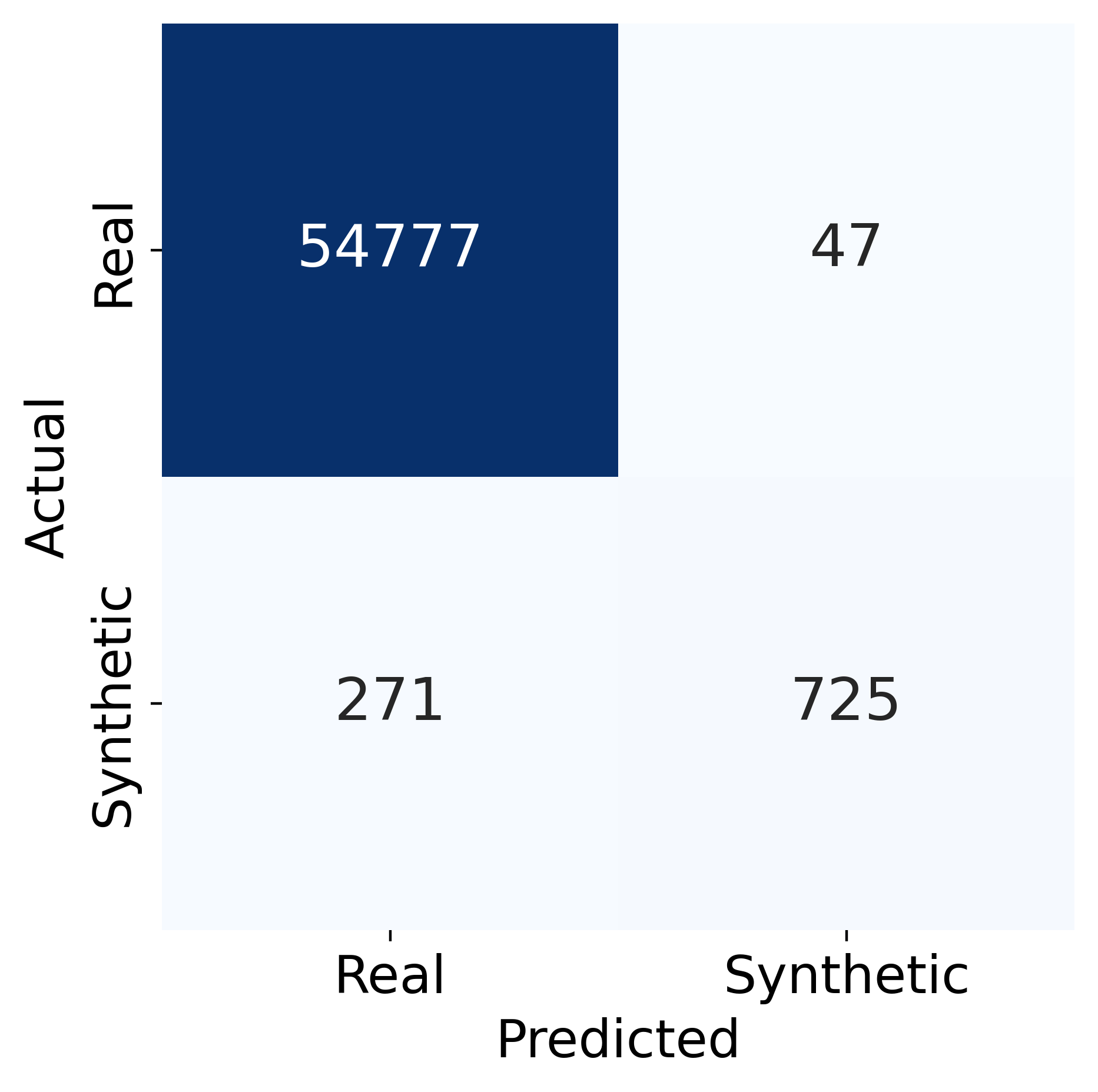} 
        \subcaption{Sequence Length 1}
    \end{minipage}
    \begin{minipage}{0.3\textwidth}
        \centering
        \includegraphics[width=0.8\textwidth]{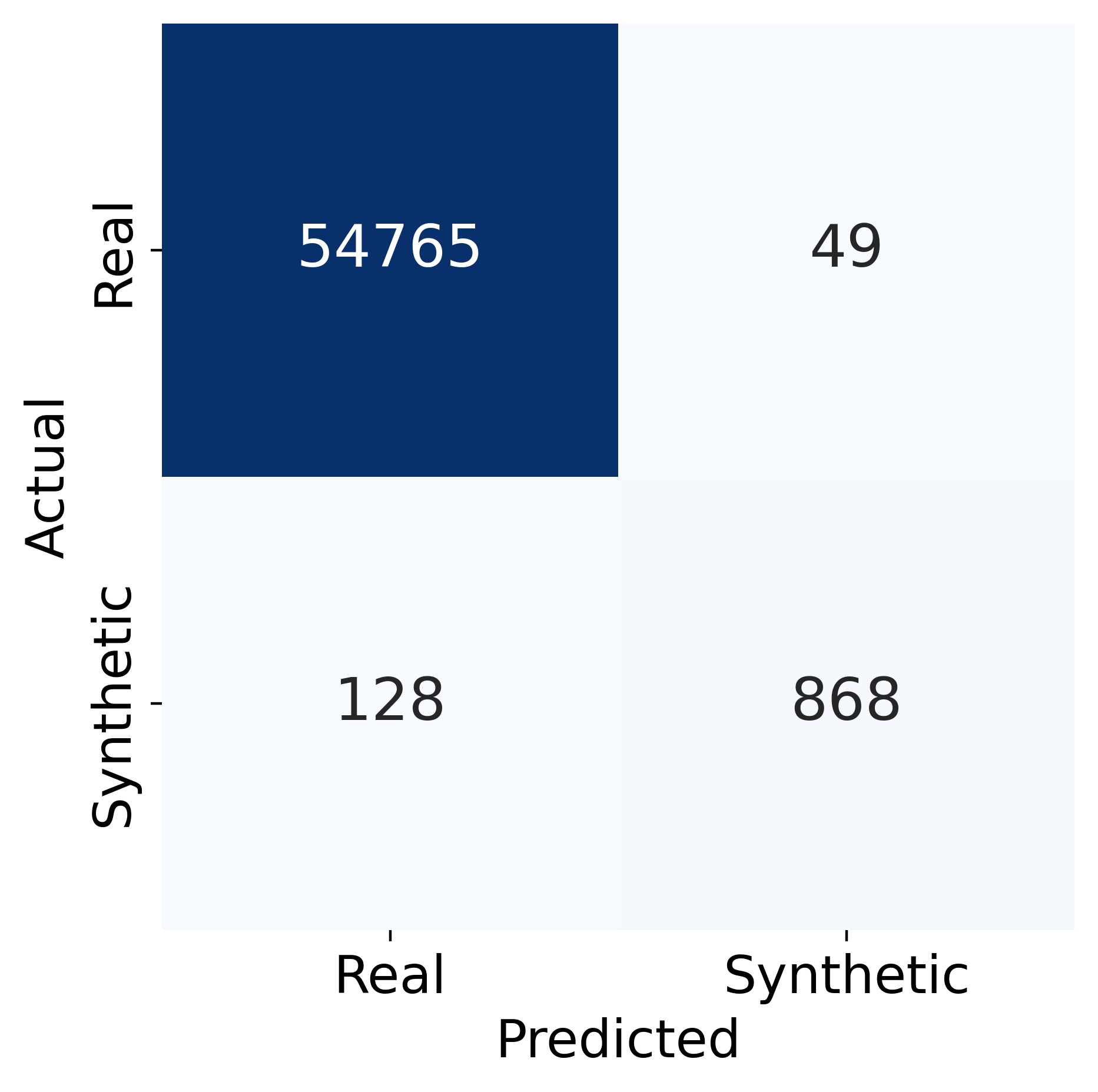} 
        \subcaption{Sequence Length 2}
    \end{minipage}
    \begin{minipage}{0.3\textwidth}
        \centering
        \includegraphics[width=0.8\textwidth]{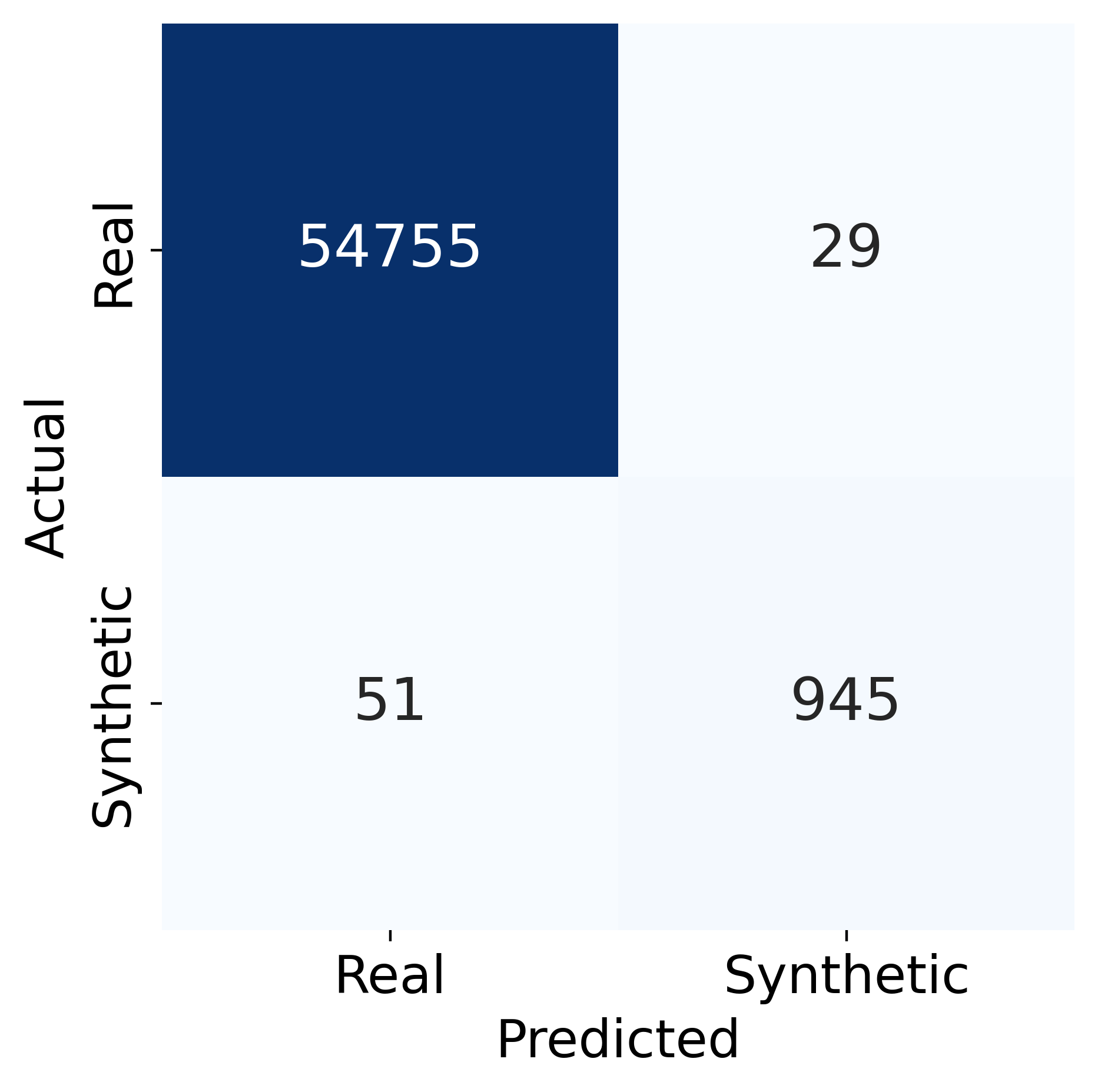} 
        \subcaption{Sequence Length 5}
    \end{minipage}
    
    \vspace{0.2cm} 
    
    \begin{minipage}{0.3\textwidth}
        \centering
        \includegraphics[width=0.8\textwidth]{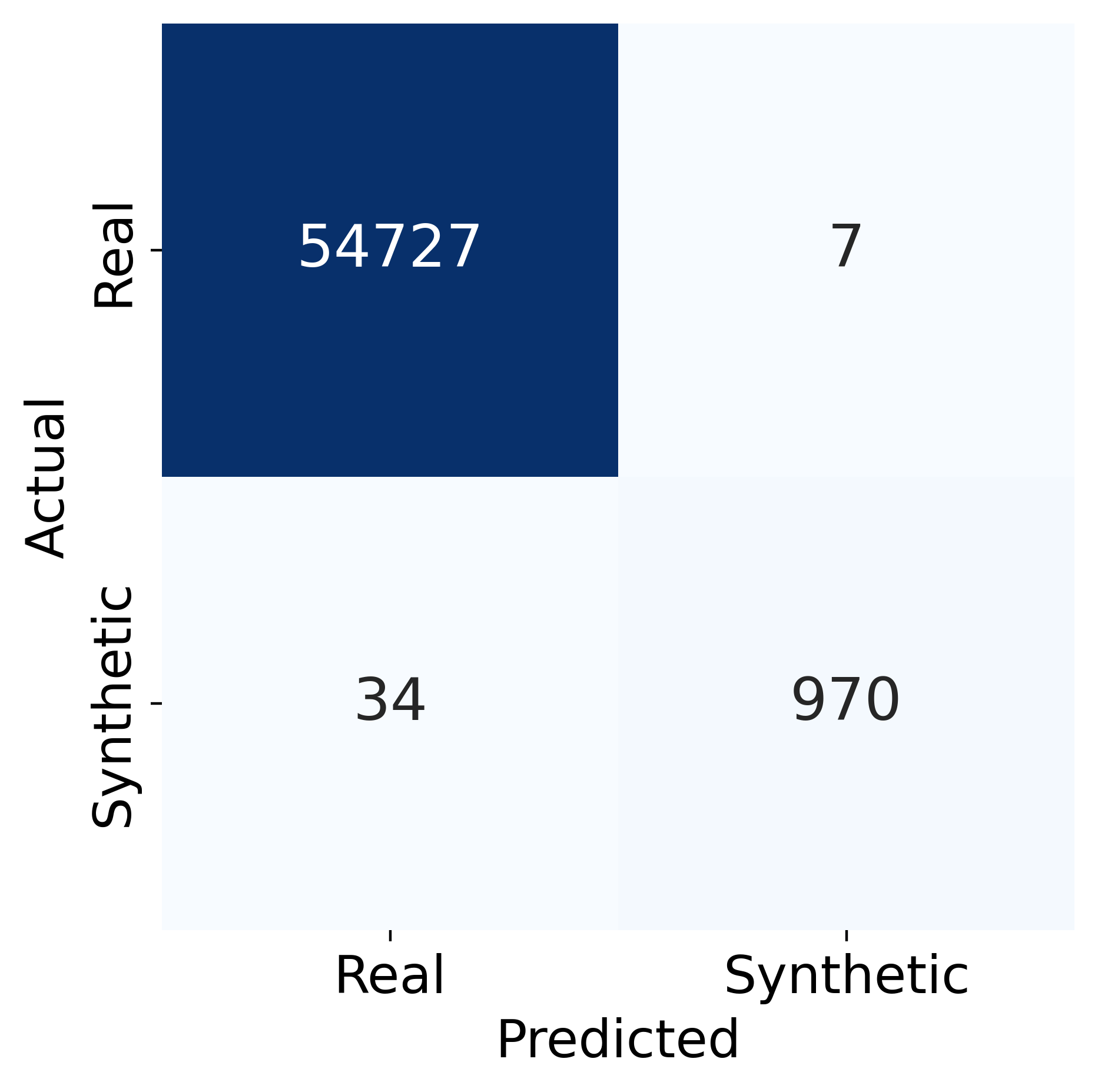} 
        \subcaption{Sequence Length 10}
    \end{minipage}
    \begin{minipage}{0.3\textwidth}
        \centering
        \includegraphics[width=0.8\textwidth]{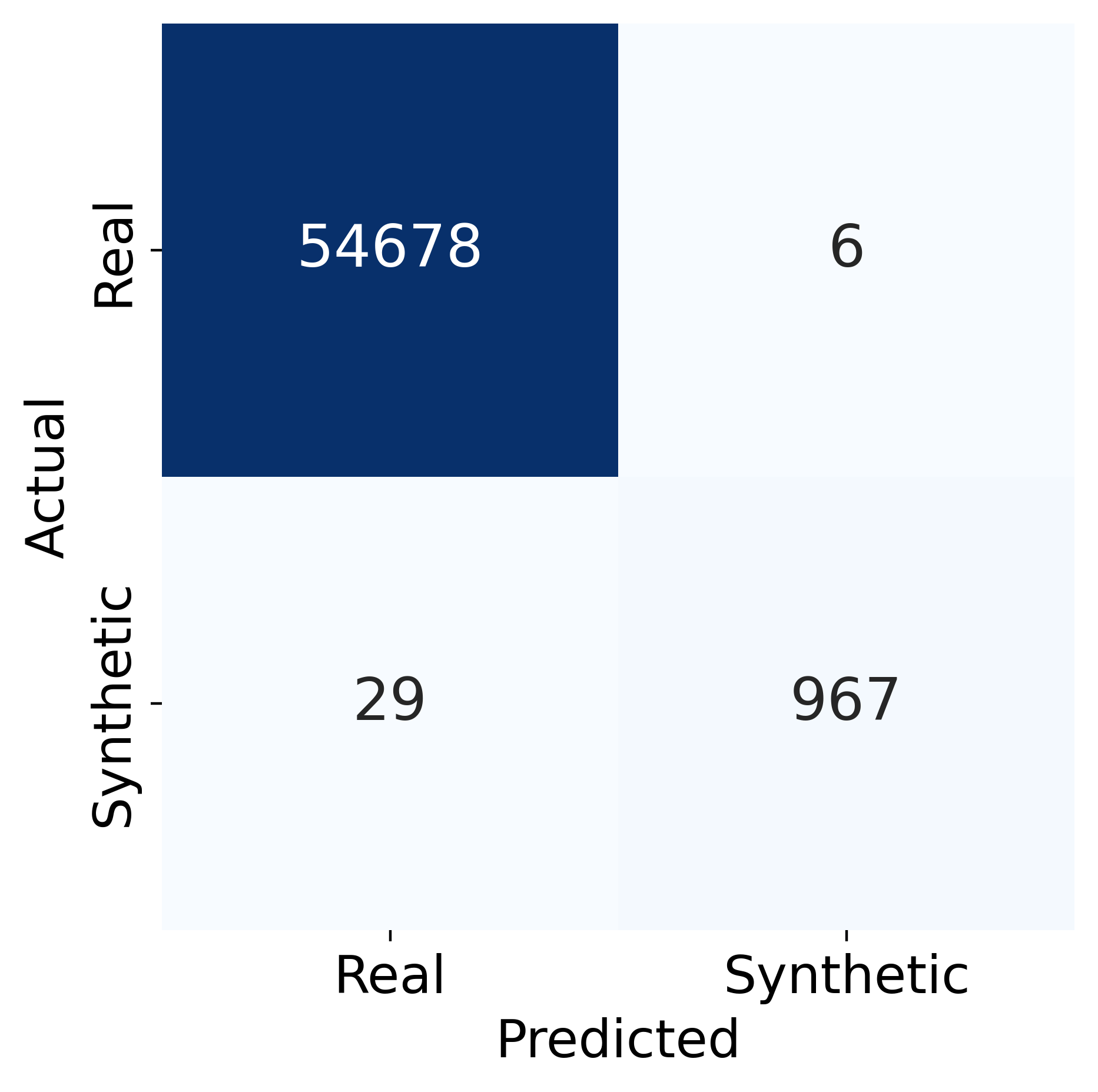} 
        \subcaption{Sequence Length 15}
    \end{minipage}
    \begin{minipage}{0.3\textwidth}
        \centering
        \includegraphics[width=0.8\textwidth]{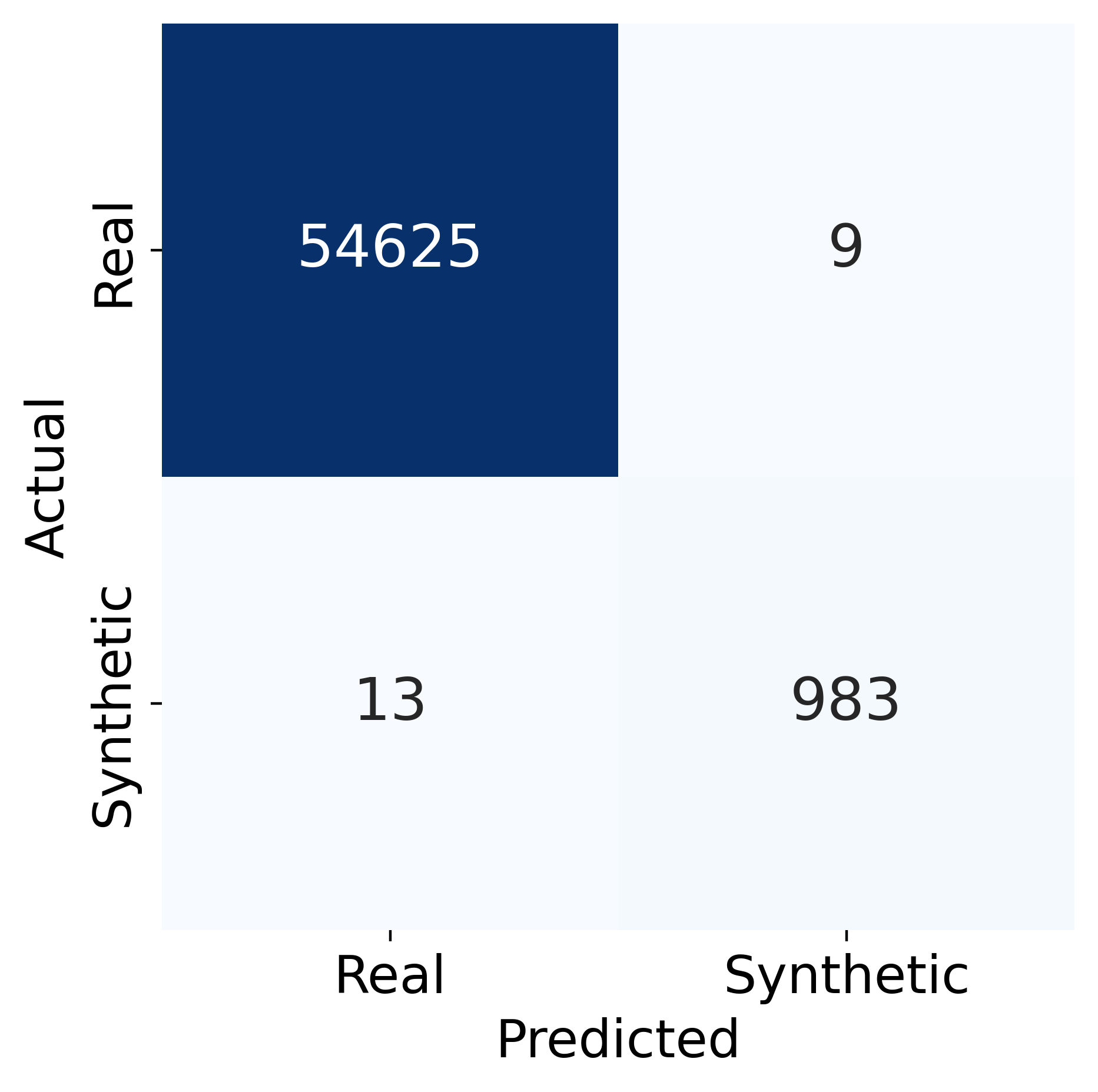} 
        \subcaption{Sequence Length 20}
    \end{minipage}
    \caption{\small Confusion Matrices for Amplification Factor 1.2}
    \label{fig:cm}
\end{figure*}

In addition, Fig~\ref{fig:cm} presents the confusion matrices for the amplification factor 1.2. This particular factor was chosen because it represents a lower level of deviation from normal data, making it more difficult for the system to predict the synthetic data accurately. Finally, the proposed detection system requires between 40 to 50 milliseconds to process and identify anomalies depending on the input size of the ML detection model. Given that the Near-RT RIC control loop requirements range from 10 to 1000 milliseconds, this detection time falls well within the acceptable range, ensuring that the system can operate efficiently within the near real-time constraints of the control loops.

\section{Conclusions and Future Directions} \label{conclusion}

In this paper, we introduced a threat in Near-RT RIC control loops of Open RAN, called KPI Poisoning Attack, that exploits vulnerabilities to manipulate the KPI values maliciously. To give a deep insight into the significant effects of the attack on the network performance and even its operation, the role of KPIs and the potential impact of falsified values of KPIs were analysed in different use cases in Near-RT RIC control loops. Based on the analysis, the attack can severely affect traffic steering and energy-saving use cases. To address this threat, we proposed a novel LSTM-based approach, which includes a procedure to check the KPI reports for anomalies before being used in control loops' decision-making. We conducted an emulation to generate KPI reports and injected anomalies into the values. Our approach effectively detected these anomalies in different conditions, as demonstrated by our experiments. The future direction for this research is to explore diverse scenarios for injecting anomalies with different consequences in network control procedures and designing an effective AI/ML-based detection/mitigation scheme. 
\section*{Acknowledgment}
This work was supported in part by the Highly Intelligent, Highly Performing RAN (HiPer-RAN) Project and in part by the Mobile oRAN for highly Dense Environments (5G MoDE) Project, Two major winners of the U.K.’s Department of Science, Innovation and Technology (DSIT) Open Networks Ecosystem Competition.

\FloatBarrier

\bibliographystyle{IEEEtran}

\end{document}